\documentclass[prl,aps,twocolumn,showpacs,citeautoscript,reprint,superscriptaddress]{revtex4-2}
\usepackage{graphicx}
\usepackage{bm,amsmath,amssymb,wasysym,amsfonts,dcolumn}
\usepackage[colorlinks=true, urlcolor=blue, linkcolor=blue, citecolor=blue, pdfborder={0 0 0}]{hyperref}
\usepackage[usenames,dvipsnames]{xcolor}
\usepackage{mathrsfs}
\usepackage{booktabs}
\usepackage{multirow}
\usepackage{makecell}

\usepackage{hyperref}
\newcommand{\DOI}[1]{\href{https://doi.org/#1}}

\begin{document}

\title{Symmetry-forbidden intraband transitions leading to ultralow Gilbert damping in van der Waals ferromagnets}

\author{Weizhao Chen}
\affiliation{Center for Advanced Quantum Studies and School of Physics and Astronomy, Beijing Normal University, Beijing 100875, China}
\affiliation{Interdisciplinary Center for Theoretical Physics and Information Sciences, Fudan University, Shanghai 200433, China}
\author{Yu Zhang}
\affiliation{Center for Advanced Quantum Studies and School of Physics and Astronomy, Beijing Normal University, Beijing 100875, China}
\author{Yi Liu}
\email{yiliu42@shu.edu.cn}
\affiliation{Institute for Quantum Science and Technology, Shanghai University, Shanghai 200444, China}
\affiliation{Department of Physics, Shanghai University, Shanghai 200444, China}
\author{Zhe Yuan}
\email{yuanz@fudan.edu.cn}
\affiliation{Interdisciplinary Center for Theoretical Physics and Information Sciences, Fudan University, Shanghai 200433, China}
\affiliation{State Key Laboratory of Surface Physics, Fudan University, Shanghai 200433, China}

\date{\today}

\begin{abstract}
Based upon first-principles calculations, we report ultralow Gilbert damping in two-dimensional (2D) van der Waals (vdW) ferromagnets. The low damping occurs at weak scattering because mirror symmetry prohibits intraband transitions. The monotonic dependence on the electronic scattering rate suggests the absent lower limit, in contrast to conventional ferromagnetic materials. Breaking mirror symmetry through magnetization rotation, layer stacking, or structural phase transition significantly increases damping by enabling intraband transitions. Topological nodal lines, also protected by mirror symmetry, contribute substantially to interband-transition-mediated damping, which can be tuned by adjusting the Fermi level. Our findings elucidate the unique characteristics of Gilbert damping in 2D vdW ferromagnets, providing valuable insights for designing low-dimensional spintronic devices with high energy efficiency.
\end{abstract}

\maketitle

{\color{red}\it Introduction.---}The discovery of 2D vdW ferromagnetic (FM) materials, exemplified by CrI$_{3}$~\cite{Huang:nat17}, Cr$_2$Ge$_2$Te$_6$~\cite{Gong:nat17}, and Fe$_{3}$GeTe$_{2}$~\cite{Fei:natm18,Deng:nat18}, has not only dramatically broadened the horizon of FM materials but also ignited promising prospects for low-dimensional spintronics devices~\cite{Gong:sc19,Dieny:nate20}. This promise stems partly from the fact that magnetism in 2D vdW FM metals can be controlled through unique approaches like gating and strain~\cite{Burch:nat18,Blei:apr21}. As a typical 2D FM metal, Fe$_{3}$GeTe$_{2}$ has attracted considerable attention for its spin-orbit torque-driven magnetization switching~\cite{Wang:sa19,Alghamdi:nl19,Wang:natcomm23}, its unconventional tunability in magnetic tunnel junctions~\cite{Wang:nl18,Wang:nl23} and its potential in novel electronic applications~\cite{Xiong:natcomm24}. A very similar vdW FM metal Fe$_{3}$GaTe$_{2}$~\cite{Zhang:natcomm22} has been discovered with an almost identical lattice structure to Fe$_{3}$GeTe$_{2}$, yet it boasts a Curie temperature above room temperature, rendering it an ideal candidate for integration into magnetic heterostructures and devices that operate under ambient conditions~\cite{Pan:acsml23,Jin:nanos23,Deng:nl24,Zhang:am24}. 

The quest for designing high-performance, low-consumption spintronics devices based on 2D vdW FM materials requires a comprehensive understanding of the dynamical properties of 2D magnetism~\cite{Tang:pr23}, particularly the effects of reduced dimensionality. Gilbert damping, a crucial parameter in magnetization dynamics~\cite{Gilbert:ieee04}, determines the critical current density for current-induced magnetization switching~\cite{Brataas:nm12}, the switching time scale~\cite{Sun:prb00} and the velocity of magnetic texture driven by magnetic fields and/or electric currents~\cite{Tatara:pr08,Wang:ne20}. More importantly, low Gilbert damping is essential for low-consumption in memory and logic devices. Ultralow damping has been observed in Fe-based alloys~\cite{Schoen:NP16,Lee:NC17,Wei:SAv21,Arora:prapplied21} and half metals~\cite{Guillemard:prapplied19,Zhang:prb20}. In FM transition metals and alloys, such as Fe, Co, Ni~\cite{Bhagat:prb74,Heinrich:jap79,Khodadadi:prl20} and FePd~\cite{Zhang:apl20,Peria:sr24}, the experimentally measured damping $\alpha$ exhibits a non-monotonic temperature dependence: $\alpha$ increases with decreasing temperature (conductivity-like) at low-temperatures and with increasing temperature (resistivity-like) at high-temperatures, limiting damping above its lower bound. However, in the recently discovered 2D FM metals, Fe$_{3}$GeTe$_{2}$ and Fe$_{3}$GaTe$_{2}$, a monotonic temperature dependence of damping has been observed in both experimental~\cite{Zhang:nl24} and theoretical studies~\cite{Yang:prb22}. Furthermore, significant damping anisotropy has been identified in these 2D FM metals~\cite{Yang:prb22,Alahmed:2dm21}, suggesting an atypical magnetization relaxation likely due to their low-dimensional lattice structures.

In this Letter, we present a theoretical calculation of the Gilbert damping in 2D vdW FM metals, using Fe$_{3}$GaTe$_{2}$ as a typical example. We demonstrate that mirror symmetry in 2D vdW systems prohibits intraband transitions, leading to the absence of conductivity-like damping. Consequently, the damping exhibits a monotonic temperature dependence without the lower limit. An ultralow $\alpha<10^{-3}$ is achieved in the single layer at low temperatures. Furthermore, symmetry-protected nodal lines near the Fermi level substantially enhance interband transitions, thereby increasing the resistivity-like damping. Our findings predict that the damping can be reduced to arbitrarily low values by decreasing the electronic scattering rate. These unique features are generally applicable to other 2D vdW materials with mirror symmetry, such as Fe$_{3}$GeTe$_{2}$ and 2H-FeTe$_2$.

{\color{red}\it Theoretical methods.---}We utilize the torque-correlation model~\cite{Kambersky:cjp70,Gilmore:prl07,Gilmore:jap08} to calculate the Gilbert damping of 2D vdW FM metals based upon first-principles electronic structure. This model captures the dynamical dissipation of magnetization due to spin-orbit coupling (SOC), corresponding to the dominant contribution to the intrinsic Gilbert damping $\alpha$~\cite{Heinrich:05}. Explicitly, $\alpha$ is given by
\begin{equation}
\begin{aligned}
\alpha = \frac{\pi g\mu_\textrm{B}}{M_s}  \sum_{n,m} \int \frac{d^{3}k}{(2 \pi)^{3}} |\Gamma^{-}_{mn}(\mathbf k)|^{2} \int d\epsilon\eta(\epsilon) A_{n\mathbf k}(\epsilon) A_{m\mathbf k}(\epsilon).\label{eq:tcm}
\end{aligned}
\end{equation}
Here $g$, $\mu_\textrm{B}$ and $M_s$ are the Land{\'e} factor, the Bohr magneton and the saturation magnetization, respectively. While indices $n$ and $m$ run over all energy bands, the energy derivative of the Fermi-Dirac distribution $\eta(\epsilon)\equiv-\partial f/\partial \epsilon$ ensures that the damping is predominantly influenced by electronic states near the Fermi surface. The matrix element $\Gamma^{-}_{mn}(\mathbf k)=\langle \Psi_{m\mathbf k} \vert [\sigma^{-}, H_\textrm{soc}] \vert \Psi_{n\mathbf k} \rangle$ characterizes the torque on electron spin stemming from SOC. The spectral function $A_{n\mathbf k}(\epsilon)$ is modeled as a Lorentzian centered at the band energy $\epsilon_{n\mathbf k}$, with its width determined by the electronic scattering rate $\gamma$~\cite{Gilmore:prl07,Gilmore:jap08}. Prior studies have shown that intraband ($n=m$) transitions yield a conductivity-like damping component, which is proportional to the electronic relaxation time, $\tau=\hbar/2\gamma$~\cite{Kambersky:cjp70,Gilmore:prl07}, while interband ($n \ne m$) transitions contribute to a resistivity-like damping component that scales monotonically with increasing $\gamma$~\cite{Kambersky:cjp76}. To explicitly evaluate these intraband and interband transitions in Eq.~(\ref{eq:tcm}), we have developed a Wannier interpolation method to perform the integral of $\vert\Gamma^-_{mn}(\mathbf k)\vert^2$ based on the first-principles electronic structures. The self-consistent band structures are calculated using Quantum ESPRESSO~\cite{Giannozzi:jpcm17}, and the Wannier functions are constructed using WANNIER90~\cite{Pizzi:jpcm20}. Here, we retain only the SOC-induced Gilbert damping that governs uniform magnetization precession, as measured by ferromagnetic resonance (FMR) experiments. The Stoner damping~\cite{Sayad:prl16,Loon:cp23,Szilva:rmp23} mediated by exchange interaction that contributes to nonuniform magnetization dynamics, and the two-magnon scattering~\cite{Arias:prb99,Lenz:prb06} that can be technically excluded from FMR measurements, are beyond the scope of the present work.
 
\begin{figure}[t]
\centering
\includegraphics[width=\columnwidth]{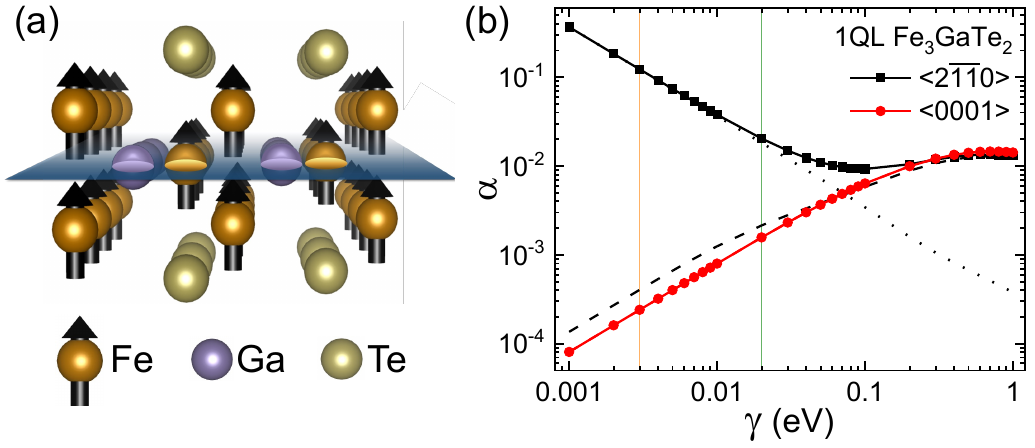}
\caption{(a) Atomic structure of 1QL Fe$_{3}$GaTe$_{2}$. The magnetic moments on the Fe atoms are indicated by black arrows. The blue plane in the middle of the layer highlights the mirror symmetry of the lattice structure. (b) Gilbert damping as a function of the electronic scattering rate for 1QL Fe$_{3}$GaTe$_{2}$ with the out-of-plane (red solid circles) and in-plane magnetization (black solid squares). The dotted and dashed lines represent the contributions from intraband and interband transitions, respectively. The orange and green vertical lines denote two characteristic scattering rates, $\gamma=3$ meV and 20 meV, which are used to calculate the damping in detail. \label{fig1}}
\end{figure}

{\color{red}\it Intraband transitions.---} We choose Fe$_3$GaTe$_2$ as a typical example to illustrate the unique characteristics of Gilbert damping in 2D vdW ferromagnets. The lattice structure of a single layer Fe$_3$GaTe$_2$, referred to as one quintuple-layer (1QL), is depicted in Fig.~\ref{fig1}(a). This lattice, belonging to the D$_{3h}$ point group, maintains mirror symmetry about the blue plane only when the magnetization is perpendicular to the atomic plane~\cite{Zhang:natcomm22}. The Gilbert damping $\alpha$ of 1QL Fe$_{3}$GaTe$_{2}$, calculated using Eq.~(\ref{eq:tcm}), is plotted in Fig.~\ref{fig1}(b) as a function of electronic scattering rate $\gamma$. With out-of-plane magnetization ($\mathbf m\|\langle0001\rangle$), $\alpha$ increases monotonically with increasing $\gamma$. Detailed analysis shows that the intraband transitions are completely absent, leading to vanishing conductivity-like behavior. As a result, $\alpha$ is lower than $10^{-3}$ at small scattering rate $\gamma<10$~meV. In contrast, for in-plane magnetization ($\mathbf m\|\langle2\bar{1}\bar{1}0\rangle$), $\alpha$ exhibits a nonmonotonic dependence on $\gamma$, with both conductivity-like and resistivity-like terms contributing.

\begin{figure}[t]
\centering
\includegraphics[width=\columnwidth]{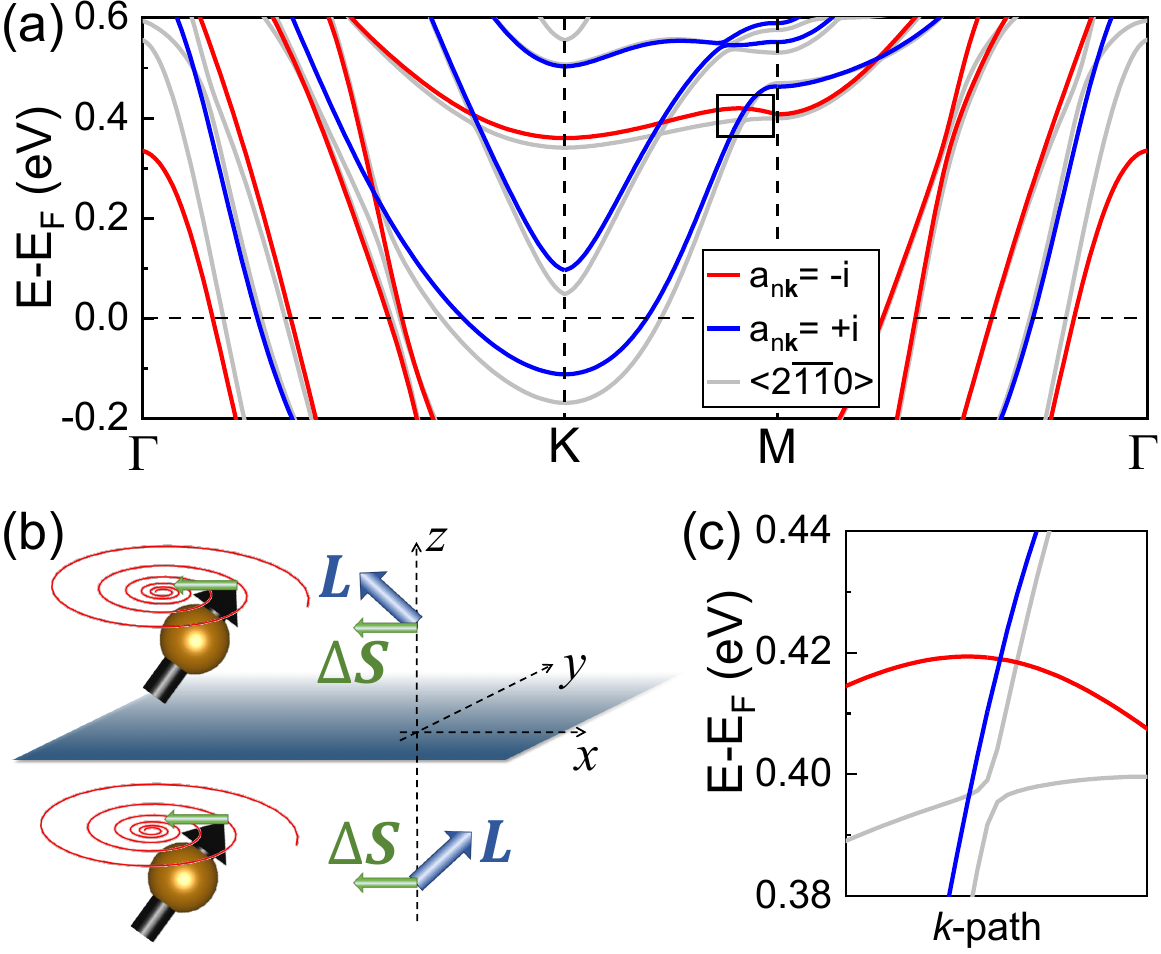}
\caption{(a) Calculated band structure of 1QL Fe$_{3}$GaTe$_{2}$ with out-of-plane magnetization. The red and blue lines denote mirror-symmetric ($a_{n\mathbf k}=-i$) and antisymmetric ($a_{n\mathbf k}=+i$) bands, respectively. Bands calculated with in-plane magnetization are presented in gray. (b) Schematic illustration of energy dissipation during magnetization dynamics in a mirror-symmetric system. (c) Enlarged view of a typical symmetry-protected band crossing, as indicated by the rectangle in (a). \label{fig2}}
\end{figure}

The ultralow damping due to the absence of conductivity-like damping with out-of-plane magnetization can be explained through a symmetry analysis of the electronic states. We define $\mathcal M_z$ as the mirror operator~\cite{Zhang:prl13} about the central plane of the 1QL Fe$_{3}$GaTe$_{2}$. For $\mathbf m\|\langle0001\rangle$, the system is mirror-symmetric, and thus $\mathcal M_z$ commutes with the total Hamiltonian, as well as the SOC Hamiltonian $H_\textrm{soc}$~\cite{Zhang:prl13}. Consequently, each Bloch state $\Psi_{n\mathbf k}$ of the system is an eigenfunction of $\mathcal {M}_{z}$ with eigenvalues $a_{n\mathbf k}=-i$ (symmetric) or $+i$ (antisymmetric). These mirror symmetry properties of Bloch states for 1QL Fe$_3$GaTe$_2$ are explicitly calculated using the IRVSP code~\cite{Gao:cpc21}, as shown in Fig.~\ref{fig2}(a), where red and blue lines represent symmetric and antisymmetric energy bands with out-of-plane magnetization. Since $\mathcal M_z$ anti-commutes with $\sigma^-$, we have $\{\mathcal M_z,\,[\sigma^-,H_{\rm soc}]\}=0$. Considering the mirror symmetry properties of Bloch states, we find
\begin{eqnarray}
 \Gamma^{-}_{mn}(\mathbf k)=-a^{*}_{m\mathbf k} a_{n\mathbf k} \Gamma^{-}_{mn}(\mathbf k).\label{eq:gam}
\end{eqnarray} 
For intraband transitions ($m=n$), $\vert a_{n\mathbf k}\vert^{2}=1$ always holds, leading to $\Gamma^{-}_{nn}(\mathbf k)=-\Gamma^{-}_{nn}(\mathbf k)$. Thus we have $\Gamma^{-}_{nn}(\mathbf k)=0$ indicating that the intraband transitions are completely forbidden, regardless of the symmetry of the Bloch states. Consequently, the conductivity-like damping vanishes and $\alpha$ increases monotonically with $\gamma$.
   
Alternatively, the symmetry-forbidden intraband transitions can be intuitively understood by examining the energy dissipation during magnetization dynamics within Kambersk{\'y}'s breathing Fermi surface model~\cite{Kambersky:cjp70,Kambersky:cjp76}. When the magnetization is out-of-plane at equilibrium, the transverse (in-plane) components of the spin angular momentum $\Delta\mathbf S$ relax, as illustrated in Fig.~\ref{fig2}(b). The energy changes with the spin variation $\Delta\mathbf S$ through SOC, i.e., $\Delta E=\Delta H_{\rm soc}\propto \mathbf L\cdot\Delta\mathbf S$. Due to the mirror symmetry, the orbital angular momentum $\mathbf L$ must have the same out-of-plane component but opposite in-plane components on either side of the mirror plane, namely $L_{z}(z)=L_{z}(-z)$ and $L_{x,y}(z)=-L_{x,y}(-z)$. Therefore, after integrating over the entire space, the band energy remains unchanged with the in-plane $\Delta \mathbf S$, i.e., $\Delta E=0$. This implies that the breathing Fermi surface during the magnetization dynamics does not excite the system out of equilibrium, and the intraband transitions are unnecessary to bring it back to equilibrium.

\begin{figure}[t]
\centering
\includegraphics[width=\columnwidth]{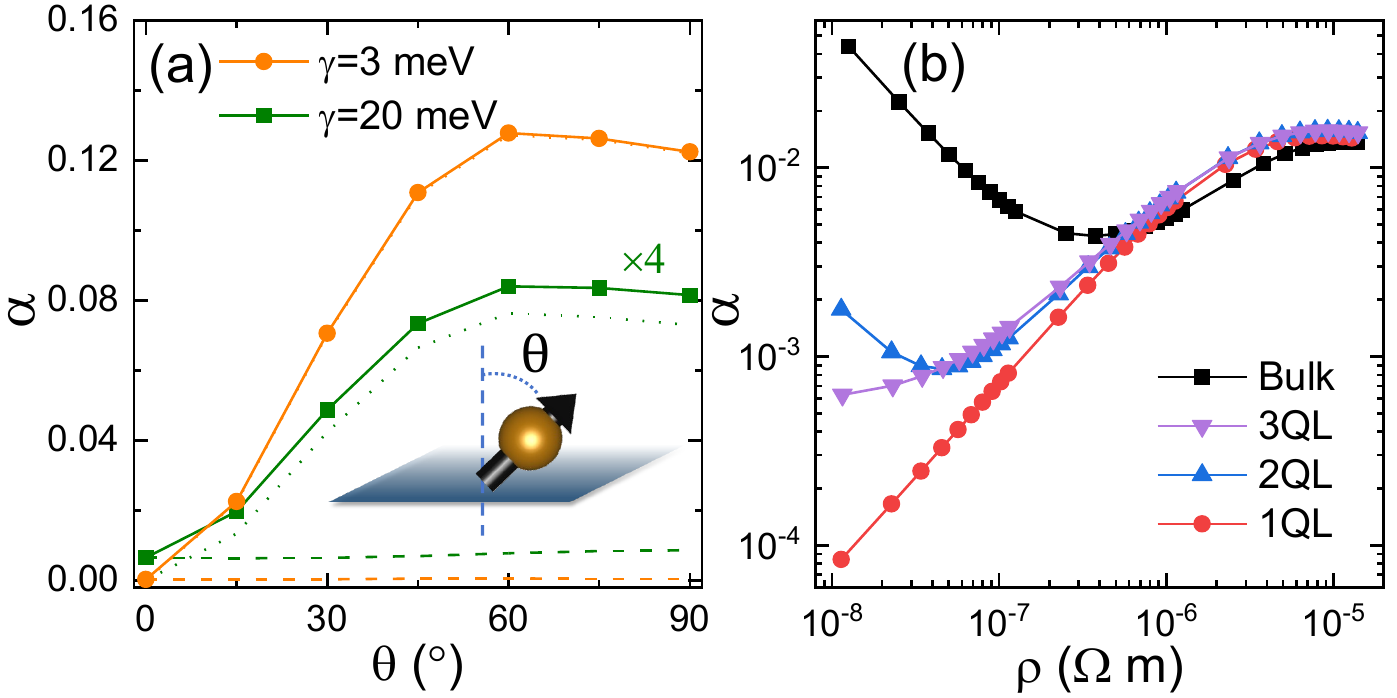}%
\caption{(a) Gilbert damping for 1QL Fe$_{3}$GaTe$_{2}$ as a function of the tilting angle $\theta$ of the magnetization away from the surface normal, for two scattering rates $\gamma=3$~meV and 20~meV. The intraband and interband contributions to the damping are represented by dotted and dashed curves, respectively. (b) Gilbert damping for Fe$_{3}$GaTe$_{2}$ with varying stacking layers plotted as a function of electrical resistivity. \label{fig3}}
\end{figure}

To further confirm that the absent intraband transitions (conductivity-like damping) is indeed a result of symmetry, we intentionally break the mirror symmetry in different ways. First, we rotate the magnetization of 1QL Fe$_{3}$GaTe$_{2}$ away from the surface normal. This rotation means the Bloch states are no longer eigenstates of $\mathcal M_z$. The calculated energy bands for in-plane magnetization are plotted with gray lines in Fig.~\ref{fig2}(a), where the hybridization of the red and blue bands forms anticrossing gaps with a typical example magnified in Fig.~\ref{fig2}(c). Such band hybridization facilitates intraband transitions, leading to the emergence of the conductivity-like damping due to intraband transitions, corresponding to the black squares in Fig.~\ref{fig1}(b). We select two scattering rates, $\gamma=3$~meV and 20~meV, as marked by the vertical lines in Fig.~\ref{fig1}(b), and calculate damping as a function of the tilting angle $\theta$. As shown in Fig.~\ref{fig3}(a), the calculated $\alpha$ increases monotonically with $\theta$ and tends to saturate at $\theta>60^\circ$. The saturation suggests that the energy bands hybridization is strong enough to no longer be treated as a perturbation. The enhanced damping is dominated by the intraband transitions (the dotted lines) for both $\gamma$ values, while the interband contributions (the dashed lines) remain nearly unchanged. We conducted an analysis on Fe$_{3}$GeTe$_{2}$, whose lattice structure is identical to Fe$_{3}$GaTe$_{2}$, except for the substitution of Ga atoms with Ge. For 1 QL Fe$_{3}$GeTe$_{2}$ with perpendicular magnetization, conductivity-like damping vanishes, resulting in a monotonic increase of the calculated $\alpha$ with increasing $\gamma$. Rotating the magnetization leads to recovery of intraband contributions, thus yielding non-monotonic behavior, which is qualitatively the same as Fe$_{3}$GaTe$_{2}$; see Sec.~1 in Supplemental Material~\cite{SM}.

The second approach involves varying the number of stacking layers of Fe$_{3}$GaTe$_{2}$. By analyzing the lattice structures of 1QL, 2QL and 3QL Fe$_{3}$GaTe$_{2}$, we find that mirror symmetry is preserved for odd-number stacking layers but broken in the 2QL system. The calculated damping of Fe$_{3}$GaTe$_{2}$ with different numbers of stacking layers is shown in Fig.~\ref{fig3}(b), plotted as a function of the simultaneously calculated electrical resistivity $\rho$. The resistivity is calculated using the Boltzmann equation within the relaxation time approximation, where $\rho$ is proportional to the scattering rate $\gamma=\hbar/2\tau$. The calculated $\alpha$ for 1QL and 3QL both monotonically increase with $\rho$, indicating the forbidden intraband transitions, as expected due to mirror symmetry. In contrast, $\alpha$ of 2QL Fe$_{3}$GaTe$_{2}$ exhibits a nonmonotonic dependence on $\rho$ with noticeable conductivity-like damping arising from intraband transitions at low $\rho$. The calculated $\alpha$ for bulk Fe$_{3}$GaTe$_{2}$, shown as the black squares in Fig.~\ref{fig3}(b), contains a much larger conductivity-like component. This is because the mirror symmetry is well defined only for the Bloch states with $k_z=0$ and those $k_z$ located at the boundary plane of the first Brillouin zone~\cite{Zhang:prl13}. Other electronic states between these special planes are not eigenstates of the mirror operator and intraband transitions are not prohibited by mirror symmetry. Compared with the bulk case, the significantly reduced damping in Fe$_{3}$GaTe$_{2}$ thin films at low $\rho$ manifests the strong suppression of the intraband transitions and thus the conductivity-like behavior. 

We further performed calculations for 2D monolayer FeTe$_{2}$ in two distinct structural phases: the trigonal prismatic 2H phase and the octahedral 1T phase. The mirror-symmetric 2H-FeTe$_{2}$ with out-of-plane magnetization exhibits monotonic damping. In contrast, the damping in 1T-FeTe$_2$, which lacks mirror symmetry, behaves non-monotonically and is independent of magnetization orientation; see Sec.~2 in Supplemental Material~\cite{SM}. These additional calculations indicate that the symmetry-forbidden intraband transition is a general feature of any mirror-symmetric 2D vdW ferromagnets.

\begin{figure}[t]
\centering
\includegraphics[width=\columnwidth]{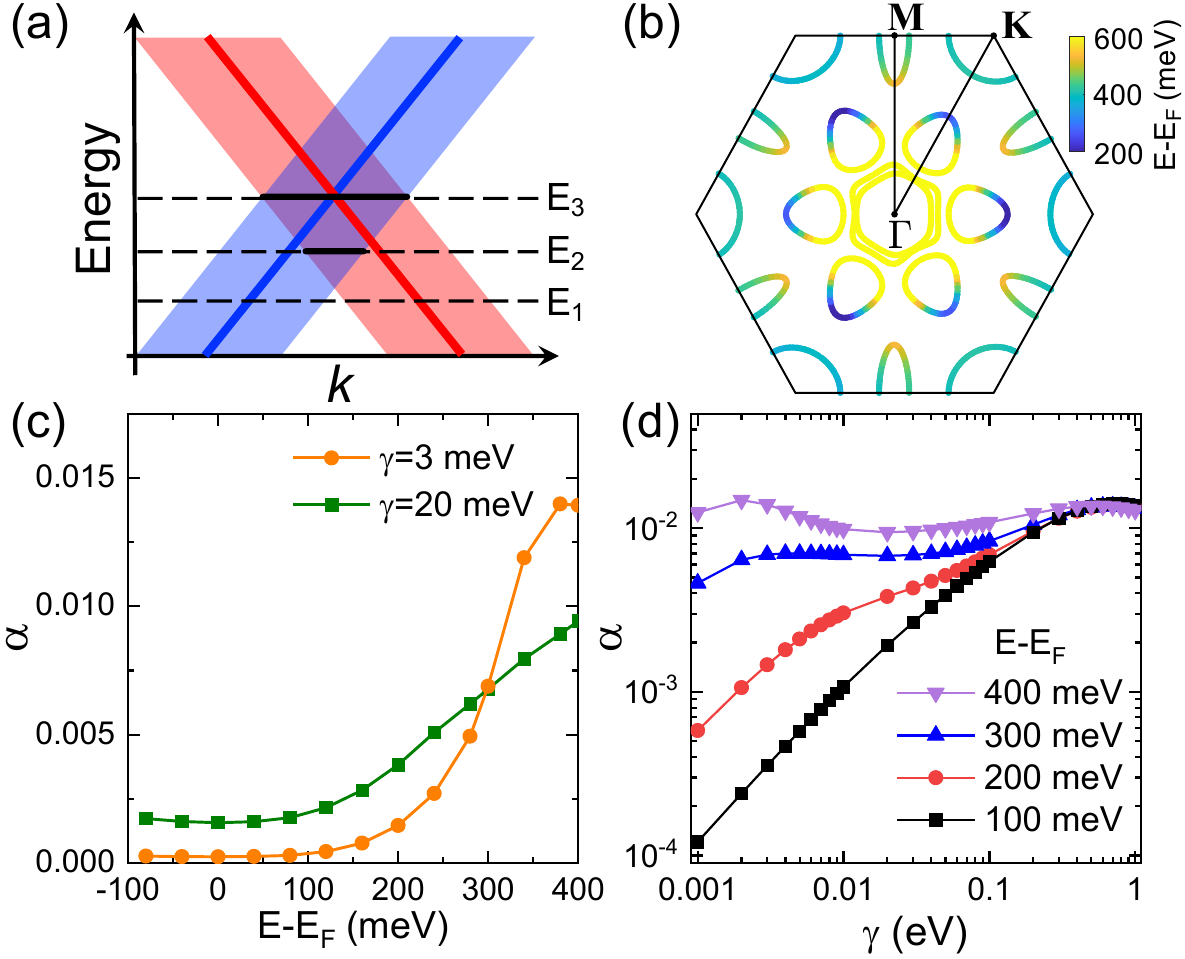}
\caption{(a) Schematic representation of band overlap near their crossing point, with shadows indicating the broadening. Thick black lines show the extent of overlap at various isoenergetic surfaces. (b) Calculated nodal lines for 1QL Fe$_{3}$GaTe$_{2}$ with out-of-plane magnetization in the 2D Brillouin zone. (c) Calculated $\alpha$ as a function of energy at two scattering rates. (d) Calculated $\alpha$ as a function of scattering rate at various energies. \label{fig4}}
\end{figure}

{\color{red}\it Interband transitions.---}Although intraband transitions are prohibited by mirror symmetry, Eq.~(\ref{eq:gam}) indicates that interband transitions are allowed between Bloch states with contrasting symmetry characteristics, i.e., $a_{m\mathbf k} \neq a_{n\mathbf k}$. In fact, the mirror symmetry ensures the presence of nodal lines in the Brillouin zone, which are formed by the intersections of mirror-symmetric and mirror-antisymmetric bands~\cite{Yang:apx18,Zeng:prl20}. According to Eq.~(\ref{eq:tcm}), the interband contribution to damping depends on the overlap of the broadened energy bands at the Fermi energy $E_F$. As schematically illustrated in Fig.~\ref{fig4}(a), the interband contribution from the two intersecting bands is maximized when the intersection point aligns with the Fermi level, i.e., $E_F=E_3$. Consequently, pronounced interband transitions are anticipated near band crossings or nodal lines close to $E_F$. In the case of 1QL Fe$_{3}$GaTe$_{2}$, the nodal lines in the 2D Brillouin zone are located several hundred meV above $E_F$, as shown in Fig.~\ref{fig4}(b). This accounts for the very low interband damping of 1QL Fe$_{3}$GaTe$_{2}$ with out-of-plane magnetization, particularly at low $\gamma$. 

The enhancement of damping by nodal lines is confirmed by elevating $E_F$ in 1QL Fe$_{3}$GaTe$_{2}$. Two scattering rates, $\gamma=3$~meV and 20~meV, were selected, and the calculated damping is plotted in Fig.~\ref{fig4}(c) as a function of energy. Near the original $E_F$, $\alpha$ is minimal and insensitive to energy shifts since the intraband transitions are forbidden and the nodal lines are far above $E_F$. For $E-E_F>200$~meV, as energy approaches the nodal lines, $\alpha$ increases rapidly for both $\gamma$ values due to significantly enhanced interband transitions. We further investigate the scattering rate dependence of $\alpha$ at various energies. As shown in Fig.~\ref{fig4}(d), the typical monotonic resistivity-like damping is observed at $E-E_F=100$ and $200$ meV, where the nodal lines are not significantly involved. When scattering occurs at $E-E_F=300$ and $400$ meV, damping becomes substantial even at very low $\gamma$. This is because the interband transitions always occur at the crossing points, irrespective of broadening. Thus, the damping displays a very weak dependence on $\gamma$. It also implies that the common assumption $\alpha_{\rm interband}\propto\rho$ does not apply in these instances.

The interband transitions also explain the higher $\alpha$ calculated for 3QL Fe$_{3}$GaTe$_{2}$ compared to 1QL, as shown in Fig \ref{fig3}(b). With mirror symmetry maintained, we analyze the band structure for 3QL (Sec.~3 in Supplemental Material~\cite{SM}). The tripling of atoms in the unit cell results in a greater number of bands in 3QL compared to 1QL. Consequently, the denser bands lead to a larger band overlap under the same broadening $\gamma$, culminating in a substantially larger interband contribution to damping for 3QL than for 1QL.

\begin{figure}[t]
\centering
\includegraphics[width=0.8\columnwidth]{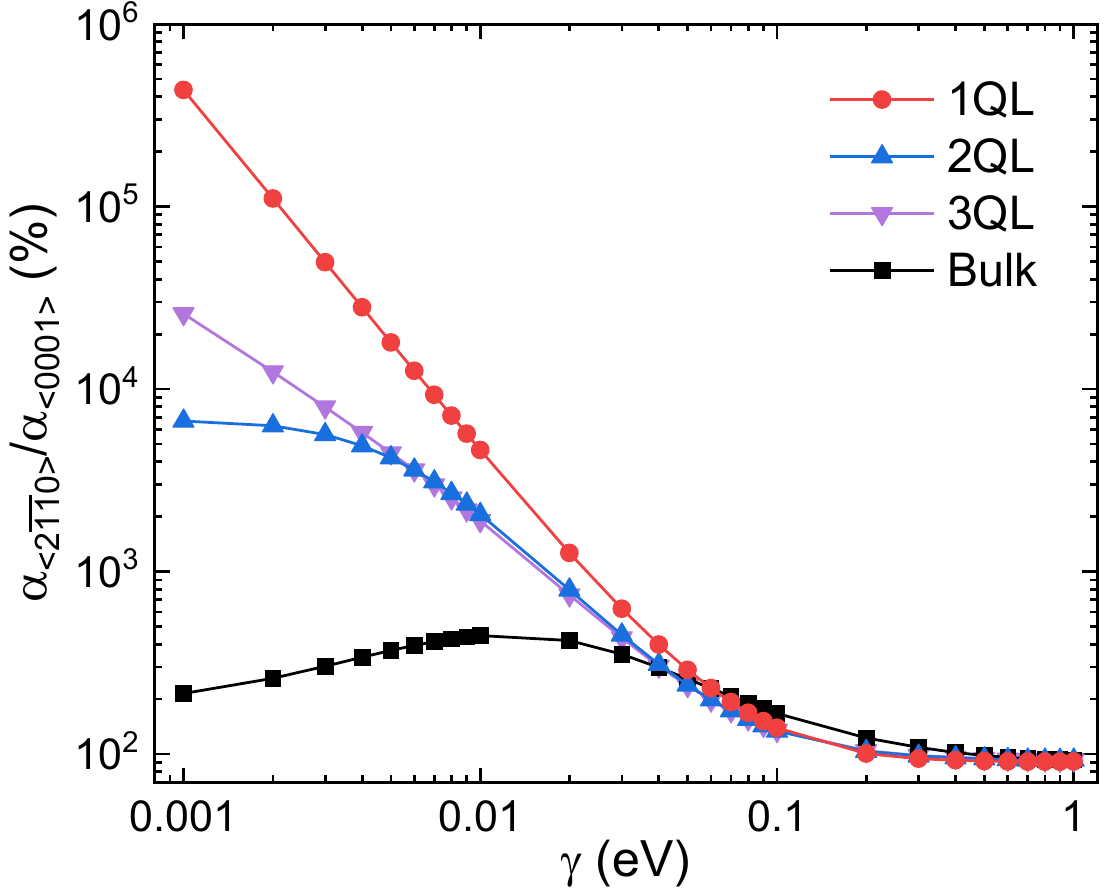}%
\caption{Calculated damping anisotropy between out-of-plane and in-plane magnetization for Fe$_{3}$GaTe$_{2}$. \label{fig5}} 
\end{figure}

{\color{red}\it Discussions and conclusions.---}The predicted features in Gilbert damping of mirror-symmetric vdW ferromagnets can be directly measured in experiments. The symmetry-forbidden intraband transitions with out-of-plane magnetization results in a low damping. By rotating the magnetization using an external field, the mirror symmetry is broken, leading to the recovery of the intraband (conductivity-like) damping. Consequently, the Gilbert damping can be controlled by adjusting the magnetization orientation without altering the scattering rate, which is particularly significant in clean systems. The calculated damping anisotropy between out-of-plane and in-plane magnetization for Fe$_{3}$GaTe$_{2}$ is depicted in Fig.~\ref{fig5}. For both 1QL and 3QL, a substantial anisotropy ratio exceeding $10^4\%$ is observed at small $\gamma$, and it exceeds 200\% for bulk. This damping anisotropy is expected to be observable using FMR or time-resolved magneto-optical Kerr effect. Additionally, the monotonic (non-monotonic) temperature-dependent damping for odd (even) layers can also be experimentally measured, as suggested by Fig.~\ref{fig3}(b).

In summary, our investigation into the Gilbert damping of 2D vdW FM metals has uncovered the crucial role of mirror symmetry in governing damping behavior. In 1QL Fe$_{3}$GaTe$_{2}$, the conductivity-like damping vanishes under out-of-plane magnetization due to symmetry-forbidden intraband transitions. This results in an ultralow Gilbert damping at small scattering rate. However, this symmetry can be broken either by tilting the magnetization or by varying the number of layers, which reinstates the intraband contribution and substantially enhances damping. Furthermore, symmetry-protected crossings between mirror-symmetric and antisymmetric bands create nodal lines where the interband transitions are pronounced, leading to a significant damping enhancement. Voltage control offers a promising avenue for tuning the damping by shifting the Fermi level relative to these nodal lines. These nodal lines and associated topological electronic states may have interesting interactions with topological magnons in vdW FM systems~\cite{Chen:prx18,Aguilera:prb20,Jaeschke-Ubiergo:prb21,Shen:prb21}. Our findings not only elucidate the microscopic origins of the observed monotonic temperature dependence of Gilbert damping but also predict a universal mechanism to achieve the ultralow Gilbert damping in 2D vdW FM materials, with no lower limit.

\acknowledgements
This work was supported by the National Key Research and Development Program of China (2024YFA1408500), the National Natural Science Foundation of China (Grants No. 12374101 and No. 12574115), the Open Fund of the State Key Laboratory of Spintronics Devices and Technologies (Grant No. SPL-2408) and the Open Project of Guangdong Provincial Key Laboratory of Magnetoelectric Physics and Devices (2022B1212010008).

\end{document}